# RDD-Eclat: Approaches to Parallelize Eclat Algorithm on Spark RDD Framework

Pankaj Singh[1], Sudhakar Singh[2,*], P. K. Mishra[1], and Rakhi Garg[3]

[1] Department of Computer Science, Banaras Hindu University, Varanasi, India
`{psingh.edu,mishra}@bhu.ac.in`
[2] Department of Electronics and Communication, University of Allahabad, Allahabad, India
`sudhakar@allduniv.ac.in`
[3] Mahila Maha Vidyalaya, Banaras Hindu University, Varanasi, India
`rgarg@bhu.ac.in`

**Abstract.** Initially, a number of frequent itemset mining (FIM) algorithms have been designed on the Hadoop MapReduce, a distributed big data processing framework. But, due to heavy disk I/O, MapReduce is found to be inefficient for such highly iterative algorithms. Therefore, Spark, a more efficient distributed data processing framework, has been developed with in-memory computation and resilient distributed dataset (RDD) features to support the iterative algorithms. On the Spark RDD framework, Apriori and FP-Growth based FIM algorithms have been designed, but Eclat-based algorithm has not been explored yet. In this paper, RDD-Eclat, a parallel Eclat algorithm on the Spark RDD framework is proposed with its five variants. The proposed algorithms are evaluated on the various benchmark datasets, which shows that RDD-Eclat outperforms the Spark-based Apriori by many times. Also, the experimental results show the scalability of the proposed algorithms on increasing the number of cores and size of the dataset.

## 1 Introduction

Frequent itemset and association rule mining [1] are the techniques of data mining employed to discover the interesting correlations among data objects of the database. These algorithms need to be re-desinged on big data processing platforms like Hadoop [2-3] and Spark [4-5] when it comes to deal with the big data. Spark is 100 times faster in memory and 10 times faster on disk than Hadoop MapReduce [4]. Many authors have designed different frequent itemset mining (FIM) algorithms on the Spark RDD framework [6-11], in which most of the algorithms follow Apriori [1] as the base algorithm. Parallelization of Eclat-based algorithm on Spark has not been explored yet to the best of our knowledge. In this paper, we consider Eclat [12], a more efficient algorithm than Apriori. Eclat reduces I/O cost due to a small number of database scan, and computation cost due to vertical data format and lattice traversal scheme.

[*]Corresponding Author

This paper proposes some approaches to parallelize Eclat algorithm on the Spark RDD framework. The name RDD-Eclat represents Spark-based Eclat algorithm, and its five variants are named in short as EclatV1, EclatV2, EclatV3, EclatV4, and EclatV5. EclatV1 is the first version of the algorithm, and each subsequent version results from the further modifications on the preceding version to achieve better performance. Algorithm EclatV1 first generate frequent items and a vertical dataset. From vertical dataset, it constructs equivalence classes based on common 1-length prefix. A default partitioner partitions the equivalence classes into *(n-1)* independent partitions, where $n$ is the number of frequent items. Equivalence classes in each partition are processed in parallel by applying the bottom-up search recursively on each equivalence class to enumerate the frequent itemsets. EclatV2 applies all operations of algorithm on the filtered transactions which contain transactions with only frequent items. Transaction filtering is adopted from the efficient implementation of Apriori and Eclat by Borgelt [13]. EclatV3 is slightly different from EclatV2, and the difference is the use of accumulator, a kind of shared variable in Spark. Algorithms EclatV4 and EclatV5 are similar to EclatV3 except the partitioner used to partition the equivalence classes. These two algorithms use two different types of hash partitioner to partition the equivalence classes into $p$ independent partitions, where $p$ is the user defined value. The performance of our proposed algorithms is compared with the Spark-based Apriori algorithm on both synthetic and real life datasets, and they significantly outperform the Spark-based Apriori in terms of execution time. Further, the performance of all proposed RDD-Eclat algorithms is compared with each other in terms of speed and scalability to study the effect of various strategies applied on these algorithms.

The rest of the paper is organized as follows. Section 2 presents preliminaries for RDD-Eclat, which is a brief description of frequent itemset mining, Eclat algorithm, and Apache Spark. Section 3 discusses the related work. In section 4, the proposed algorithms are described in detail. Experimental results and analysis are presented in section 5. Finally, section 6 concludes the paper with future directions.

## 2 Preliminaries

### 2.1 Frequent Itemset Mining and Eclat Algorithm

Frequent itemset mining is the computation of all frequent itemsets in a given database [1]. The generation of all frequent itemsets is a computationally as well as memory, and disk I/O intensive task [14]. Eclat algorithm [12] uses a vertical tidset database format, equivalence class clustering, and bottom-up lattice traversal; which reduces these costs. Eclat converts horizontal database into vertical database, i.e. from itemset format $<TID_i, i_1, i_2, ..., i_k>$ to tidset format $<i_k, TID_1, TID_2, ..., TID_k>$. In horizontal database, each transaction $T_i$ comprise of an unique transaction identifier $TID_i$ and an itemset, i.e. in the form of $<TID_i, i_1, i_2, ..., i_k>$. A vertical tidset database consists of a list of items followed by respective tidsets. The tidset of an item or itemset $X$ is the set of all transaction identifiers containing $X$, and is denoted as $tidset(X) = \{T_i.TID \mid T_i \in D, X \subseteq T_i\}$. The support of an item or itemset $X$ is the number of elements in $tidset(X)$ i.e. $\sigma(X) = |tideset(X)|$ [15]. An itemset $X$ is said to be frequent if $\sigma(X) \geq min\_sup$, where $min\_sup$ is a user-specified minimum support threshold. The tidset approach reduces the cost of support counting. The support of a candidate k-itemset is computed by the intersection of tidsets of its two (k-1)-subsets. The vertical database is more compact than horizontal and contains all relevant information, which reduces memory requirements and scanning of the whole database. Further, as the length of itemsets increases, their tidset decrease, that consequently reduces the cost of intersection operations. The computation of frequent 2-itemsets is costlier with vertical format in comparison to the horizontal format. So a triangular matrix is used to update the counts of candidate 2-itemsets [12] [14].

**Algorithm 1.** Bottom-Up recursive function of Eclat

Input: $EC_k = \{A_1, A_2, ..., A_n\}$, equivalence class of k-itemsets consists of atoms $A_i$.
Output: Frequent itemsets $\in EC_k$
1:   Bottom-Up($EC_k$)
2:   {
3:      for(i = 1; i <= |$EC_k$|; i++)
4:      {
5:         $EC_{k+1} = \phi$;
6:         for(j = i + 1; j <= |$EC_k$|; j++)
7:         {
8:            $A_{ij} = A_i \cup A_j$;
9:            tidset($A_{ij}$) = tidset($A_i$) $\cap$ tidset($A_j$);
10:           if( |tidset($A_{ij}$)| >= min_sup)
11:           {
12:              $EC_{k+1} = EC_{k+1} \cup A_{ij}$;
13:              $L_{ECk} = L_{ECk} \cup A_{ij}$;
14:           }
15:        }
16:        if($EC_{k+1}$ != $\phi$)
17:           Bottom-Up($EC_{k+1}$);
18:     }
19:     return $L_{ECk}$;
20: }

The set of items $I$ of the database forms a power-set lattice $\rho(I)$. The set of atoms of this lattice corresponds to the set of items [12]. The power-set lattice is the search space that contains all the potential frequent itemsets. To enumerate all the frequent itemsets, lattice must be traversed along with intersection operations on tidsets to obtain support count of itemsets. The equivalence class clustering partitions the lattice into smaller independent sub-lattices enabling parallel computation of frequent itemsets. It also overcomes the limited memory constraint when the complete lattice could not fit in memory due to the large intermediate tidsets. The equivalence class clustering partitions the itemsets of lattice into *equivalence classes* based on the common prefixes of itemsets. Suppose, the set of frequent k-itemsets $L_k$ is lexicographically sorted, then its itemsets can be partitioned into equivalence classes based on their common *(k-1)* length prefixes. All the classes can be processed independently and parallelly, and if a class is large enough to be solved in main memory, it can be decomposed to the next level. Eclat uses a bottom-up lattice traversal scheme [12] that processes each equivalence class by recursively decomposing into smaller classes to enumerate all frequent itemsets. The pseudo code in the Algorithm 1 shows the recursive procedure of this bottom-up search technique, originally given by Zaki [12]. Here, $L_{ECk}$ represents the set of frequent itemsets generated by the equivalence class $EC_k$. A detailed and illustrative explanation of Eclat algorithm and searching technique can be found in the base paper [12], and is not duplicated here.

**2.2   Apache Spark**

Apache Spark [4] is a fast and general cluster computing system for large-scale batch and streaming data processing, originally developed at AMPLab of UC Berkeley [5], [16]. Spark was developed to overcome the inefficiency of Hadoop MapReduce [3] [17] for iterative jobs and interactive analytics. It retains the good properties of MapReduce like scalability and fault tolerance. The backbone of Spark is a distributed memory abstraction called Resilient Distributed Datasets (RDDs) [16], which is a collection of immutable data objects partitioned across the nodes of Spark cluster. Spark achieves fault tolerance through a lineage chain that keeps record of set of dependencies on parent RDDs i.e. how an RDD derived from another RDD. A lost partition of RDD can be rebuilt quickly through the lineage chain. The more about RDD and its various operations, Spark application, and architecture of Spark cluster can be found in [5], [16], [18-19].

**3   Related Work**

With the evolution of big data, re-designing of traditional data mining algorithms on Hadoop and Spark have been started to provide the scalability. With the introduction of Hadoop, researchers have proposed several FIM algorithms on Hadoop MapReduce framework based on the central algorithms Apriori [1], Eclat [12], and FP-Growth [20]. The well known Apriori-based algorithms are SPC,

5FPC, and DPC [21]. More formal and optimized versions of these algorithms are proposed by Singh et al. [22]. Recently, Chon and Kim [23] proposed BIGMiner, an Apriori-based frequent itemset mining algorithm on MapReduce. Two distributed versions of Eclat algorithms on MapReduce have been proposed by Moens et al. [24]. The algorithms are named as Dist-Eclat and BigFIM. Dist-Eclat partitions the search space on Mappers rather than data space. Big-FIM is a hybrid of Apriori and Eclat approaches. Liu et al. [15] have incorporated three improvements in Eclat algorithm and proposed Peclat (Parallel Eclat) algorithm that parallelizes this improved algorithm on MapReduce framework. PFP (Parallel FP-Growth) [25] is a MapReduce-based FP-Growth algorithm. It breaks the FP-Tree into smaller independent FP-Trees, which are processed by different Mappers to generate frequent itemsets. FiDoop [26], a parallel frequent itemset mining algorithm on MapReduce uses an FIU-Tree (frequent items ultrametric tree) in the place of FP-Tree.

The development of Spark has shifted the research focus from Hadoop MapReduce-based algorithms to the Spark-based algorithms. MapReduce does not fit in with the iterative nature of data mining algorithms. Each time, for a new iteration, one needs to launch a new MapReduce job that takes a significant amount of time. Further, the costly read/write operations on HDFS are required for the intermediate result of jobs. Spark keeps the good features of MapReduce, resolves the problems with MapReduce, and adds a number of additional features.

During the recent years, many Spark-based FIM algorithms have been proposed. Qiu et al. [6] have proposed a Spark-based Apriori algorithm named YAFIM (Yet Another Frequent Itemset Mining). YAFIM is modularized into two phases. The first phase produces frequent items, whereas the second phase generates frequent (k+1)-itemsets from frequent k-itemsets for $k \geq 2$. YAFIM outperformed the MapReduce-based Apriori around 25 times. Rathee et al. [7] proposed R-Apriori (Reduced-Apriori), a parallel Apriori-based algorithm on the Spark RDD framework. R-Apriori is similar to YAFIM with an additional phase that reduces the computation to generate 2-itemsets. Adaptive-Miner [8] is an improvement over the R-Apriori, which dynamically selects a conventional or reduced approach of candidate generation, based on the number of frequent itemsets in recent iteration. DFIMA (Distributed Frequent Itemset Mining Algorithm) [9] is also an Apriori-based algorithm on Spark. It uses a matrix-based pruning approach to reduce the number of candidate itemsets. In the first step, it creates a Boolean vector for each frequent item, and then 2-itemset matrix from Boolean vectors. In the second step, it generates all frequent (k+1)-itemsets from frequent k-itemsets, for $k \geq 2$. HFIM (Hybrid Frequent Itemset Mining) [10] exploits the vertical format of the dataset with Apriori algorithm. The smaller size of vertical dataset reduces the cost of dataset scanning. The first phase of the algorithm produces vertical dataset that contains only frequent items. Also, a revised horizontal dataset is obtained by removing infrequent items from the original dataset. Horizontal dataset is distributed on all worker nodes while vertical dataset is shared. Shi et al. [11] proposed DFPS (Distributed FP-growth

Algorithm based on Spark) algorithm. The first step of the algorithm calculates frequent items from RDD of transactions. The second step repartitions the conditional pattern base, and the third step generates frequent itemsets in parallel from the independent partitions.

## 4 RDD-Eclat Algorithms

We parallelize Eclat algorithm on Spark RDD framework and named it as RDD-Eclat. We propose five different variants of RDD-Eclat by successively applying different strategies and heuristics. EclatV1 is the first implementation, and its successors EclatV2, EclatV3, EclatV4, and EclatV5 are resulted after applying the changes in their respective preceding algorithm. All proposed algorithms are modularized into three to four phases. Each phase comprises of transformations, actions, and other operations [19].

### 4.1 EclatV1

EclatV1 is divided into three phases described as pseudo codes in Algorithm 2, 3, and 4 respectively. Phase-1 (Algorithm 2) takes input as horizontal database and produces output as frequent items with support count, the number of frequent items, and the database in a vertical format for the subsequent use. It first creates an RDD, *transactions* from the database. Here, the database is not partitioned in order to assign a unique transaction identifier, when it is not present in the database. The *flatMapToPair()* transformation maps each transaction to a *(item, tid)* pairs, and creates a paired RDD containing the *(key, value)* pairs. The *groupByKey()* transformation groups all pairs with the same key. The *filter()* transformation removes the items having support count less than *min_sup*. The paired RDD, *freqItemCounts* contains *(item, count)* pairs, where count is the support count of item. Here, *(itemTid._1, itemTid._2)* is a *(key, value)* pair of a Tuple2 [19] type object, *itemTid*. Finally, the action, *collect()* returns the entire content of RDD, *freqItemTids* to the driver program where it is sorted in the ascending order of support and stored in a list.

**Algorithm 2.** Phase-1 of EclatV1
1:  RDD transactions = sc.textFile("database", 1);
2:  PairRDD itemTids = transactions.flatMapToPair(t -> {
3:    tid = 1;
4:    for each item of t.split(" ")
5:      pairList.add((item, tid));
6:    tid++;
7:    return pairList;
8:  }).groupByKey();
9:  PairRDD freqItemTids = itemTids.filter(itemTid -> itemTid._2.size () >= min_sup);
10: PairRDD freqItemCounts =
                 freqItemTids.mapToPair(itemTid -> (itemTid._1, itemTid._2.size()));
11: freqItemCounts.saveAsTextFile("frequentItems");
12: freqItemTidsList = sort(freqItemTids.collect());
13: n = freqItemTidsList.size();

Phase-2 of EclatV1 (Algorithm 3) computes support count of all 2-itemsets using an upper triangular matrix from the horizontal database, as recommended by Zaki in [12]. It is computed in parallel on the different partitions of the database. The database is partitioned as per the default parallelism which is equal to the number of cores on all machines of the Spark cluster. The triangular matrix is shared as an accumulator variable, *accMatrix* among all the executors to add support count of 2-itemsets in parallel. The transformation, *flatMap()* updates the accumulated matrix for all 2-itemset combination of each transaction.

**Algorithm 3.** Phase-2 of EclatV1
1:  transactions = transactions.repartition(sc.defaultParallelism());
2:  if(triMatrixMode)
3:  {
4:    create a triangular matrix, triMatrix[ ][ ]
5:    pass triMatrix as accumulator variable, accMatrix
6:    transactions.flatMap(t -> {
7:      for each 2-itemset combination, itemIitemJ of items of t.split(" ")
8:        accMatrix.update(itemIitemJ);
9:    });
10:   triMatrix = accMatrix.value();
11: }

Phase-3 of EclatV1 (Algorithm 4) takes the input as *freqItemTidsList*, the vertical dataset, and produces frequent k-itemsets, k ≥ 2. It first generates *ECList*, a list of pairs of equivalence classes for 2-itemsets and tidset of members of the class. A paired RDD, *ECs* is created by parallelizing the *ECList*, and partitioned into default *(n-1)* partitions, where *n* is the number of frequent items. The triangular matrix containing the support count of 2-itemsets is used here to avoid the costly intersection operations for infrequent 2-itemsets. The transformation, *flatMap()* processes each partition of the equivalence classes *ECs* in parallel. It applies the *Bottom-Up()* recursive function of Eclat (Algorithm 1) on each equivalence class *EC* in a partition. The source code of *Bottom-Up()* method has been taken from the SPMF Open-Source Data Mining Library [27].

---

**Algorithm 4.** Phase-3 of EclatV1
```
1:   for(i = 0; i < freqItemTidsList.size() - 1; i++)
2:   {
3:      itemI = freqItemTidsList.get(i)._1;
4:      tidsetI = freqItemTidsList.get(i)._2;
5:      for(j = i + 1; j < freqItemTidsList.size(); j++)
6:      {
7:         itemJ = freqItemTidsList.get(j)._1;
8:         if(triMatrixMode)
9:            if(triMatrix.getSupport(itemI, itemJ) < min_sup)
10:              continue;
11:        tidsetJ = freqItemTidsList.get(j)._2;
12:        tidsetIJ = tidsetI ∩ tidsetJ;
13:        prefixIList.add((itemJ, tidsetIJ));
14:     }
15:     ECList.add(itemI, prefixIList);
16:  }
17:  PairRDD ECs = sc.parallelize(ECList);
18:  ECs = ECs.partitionBy(new defaultPartitioner(n - 1)).cache();
19:  RDD freqItemsets = ECs.flatMap(EC -> Bottom-Up(EC));
20:  freqItemsets.saveAsTextFile("frequentItemsets");
```

---

### 4.2  EclatV2

EclatV2 comprises of four phases, the pseudo codes of first three phases are described in Algorithms 5, 6, and 7, whereas Phase-4 is same as the Phase-3 of EclatV1 (Algorithm 4). Phase-1 simply saves the frequent items and their support, and produces a list of frequent items in alphanumeric order.

---

**Algorithm 5.** Phase-1 of EclatV2
```
1:   RDD transactions = sc.textFile ("database");
2:   RDD items = transactions.flatMap(t -> List(t.split(" ")));
3:   PairRDD itemPairs = items.mapToPair(item -> (item, 1));
4:   PairRDD itemCounts = itemPairs.reduceByKey((v₁, v₂) -> v₁ + v₂);
5:   PairRDD freqItemCounts =
                  itemCounts.filter(itemCount -> itemCount._2 >= min_sup);
6:   freqItemCounts.saveAsTextFile("frequentItems");
7:   freqItemList = sort(freqItemCounts.keys().collect());
8:   n = freqItemList.size();
```

Phase-2 of EclatV2 (Algorithm 6) is similar to the Phase-2 of EclatV1 except the addition of transaction filtering [13]. The frequent items, *trieL$_1$* stored in a prefix tree, must be broadcasted to all executors using the broadcast variable,

4before applying the transformation. The support counting of 2-itemsets is performed applying the same triangular matrix method of EclatV1, but on the filtered transactions.

**Algorithm 6.** Phase-2 of EclatV2
```
1:   store frequent items in trie, trieL₁;
2:   RDD filteredTransactions = transactions.map(t -> filterTransaction(t.split(" "), trieL₁));
3:   if(triMatrixMode)
4:   {
5:      create a triangular matrix, triMatrix[ ][ ]
6:      pass triMatrix as accumulator variable, accMatrix
7:      filteredTransactions.flatMap(t -> {
8:         for each 2-itemset combination, itemIitemJ of items of t.split(" ")
9:            accMatrix.update(itemIitemJ);
10:     });
11:     triMatrix = accMatrix.value();
12:  }
```

Phase-3 of EclatV2 (Algorithm 7) generates the vertical dataset from filtered horizontal dataset. It first reduces all partitions of transactions into one partition in order to generate unique transaction identifier. The action *collect()* returns the list of *(item, tidset)* pairs, that is sorted by the total order of increasing support count and stored in a list. Phase-4 of EclatV2 is exactly same as the Algorithm 4, where equivalence classes are created and partitioned for parallel computation of the frequent itemsets.

**Algorithm 7.** Phase-3 of EclatV2
```
1:   filteredTransactions = filteredTransactions.coalesce(1);
2:   PairRDD freqItemTids = filteredTransactions.flatMapToPair(t -> {
3:      tid = 1;
4:      for each item of t.split(" ")
5:         pairList.add((item, tid));
6:      tid++;
7:      return pairList;
8:   }).groupByKey();
9:   freqItemTidsList = sort(freqItemTids.collect());
```

### 4.3   EclatV3

EclatV3 comprises of four phases in which first two phases, Phase-1 and Phase-2 are exactly same as those of EclatV2. The purpose of Phase-3 of both algorithms EclatV2 and EclatV3 is same i.e. both generate vertical dataset. The difference is that EclatV3 uses a hashmap data structure to store *(item, tidset)* pairs of vertical dataset. This hashmap is accumulated across all executors, and updated by the *flatMapToPair()* transformation. The updated hashmap is used to sort the list of frequent items of Phase-1, by total order of increasing support count. Phase-4 of EclatV3 is similar to the Algorithm 4, the only difference is the data structure used to store the pairs of item and tidset. Here, the items and corresponding tidsets are fetched from a hashmap instead of a list, and the rest of process is the same.

### 4.4 EclatV4 and EclatV5

Algorithms EclatV4 and EclatV5 apply the heuristics on EclatV3 to partition the equivalence classes into $p$ partitions, where $p$ has the value supplied by the user. Heuristics are applied to balance the partitions of equivalence classes. Only the Phase-4 of these two algorithms is different from EclatV3, and first three phases are the same to those of EclatV3. Further, Phase-4 is different only with respect to the partitioning of equivalence classes which is done at line no. 18 (e.g. in Algorithm 4). EclatV4 and EclatV5 respectively use *hashPartitioner* and *reverseHashPartitioner* in their Phase-4 instead of *defaultPartitioner* as in Algorithm 4.

The hash partitioner of EclatV4 applies a hash function on the values corresponding to the prefix of equivalence classes, and returns the remainder as a partition ID. Whereas, EclatV5 returns the partition ID in reverse order when the unique value assigned to the 1-length prefix of equivalence class is greater than or equal to $p$. The partitioners with hashing and reverse hashing are used to investigate the workload balance among partitions. The workload is measured in terms of the members in equivalence classes. An equivalence class having more members leads to the generation of more candidate itemsets as well as the intersection of their tidsets.

## 5 Experimental Results

The Experiments are conducted on a workstation machine installed with Spark-2.1.1, Hadoop-2.6.0, and Scala-2.11.8. The workstation is equipped with Intel Xenon CPU E5-2620@2.10 GHz with 24 cores, 16 GB memory and 1 TB disk, and running 64 bit Ubuntu 14.04. HDFS is used as storage for the input datasets and generated frequent itemsets. Source codes of all algorithms are written in Java-7. Table 1 summarizes datasets used in experiments with their properties. BMS_WebView_1 (BMS1) and BMS_WebView_2 (BMS2) are the click-stream data taken from [27] while T10I4D100K and T40I10D100K are generated by IBM Generator available at [28].

**Table 1.** Datasets used in experiments with their properties

| Dataset | Type of dataset | Transactions | Items | Average Transaction Width |
|---|---|---|---|---|
| BMS_WebView_1 | Real-life | 59602 | 497 | 2.5 |
| BMS_WebView_2 | Real-life | 77512 | 3340 | 5 |
| T10I4D100K | Synthetic | 100,000 | 870 | 10 |
| T40I10D100K | Synthetic | 100,000 | 1000 | 40 |

The proposed algorithms EclatV1, EclatV2, EclatV3, EclatV4, and EclatV5 require two parameters, *triMatrixMode* and *p* to be set before the execution. The triangular matrix optimization can be enabled or disabled by providing the true or false value to *triMatrixMode*. It is true for all datasets except BMS1 and BMS2. The size of triangular matrix depends on the maximum integer value of all items in the dataset, and it is very large in BMS1 and BMS2. The very large size of the matrix may cause out of memory problem, so the value of *triMatrixMode* is false for these two datasets. Further, algorithms EclatV4 and EclatV5 partition the equivalence classes into *p* partitions, which is set as 10 for all datasets.

### 5.1 Execution Time on Varying Value of Minimum Support

Figs. 1(a)-4(a) compare the execution time of the proposed algorithms against the Apriori algorithm whereas Figs. 1(b)-4(b) compare the execution time of the proposed algorithms EclatV1, EclatV2, EclatV3, EclatV4, and EclatV5. On all datasets, RDD-Eclat outperforms the RDD-Apriori (Figs. 1(a)-4(a)), and the execution time difference between them becomes wider with the decreasing value of minimum support.

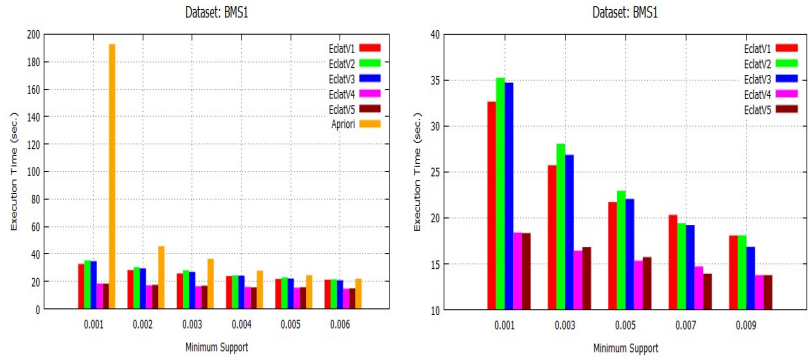

**Fig. 1.** Execution time of algorithms (a) Eclat variants and Apriori (b) Only Eclat variants for varying minimum support on dataset BMS_WebView_1

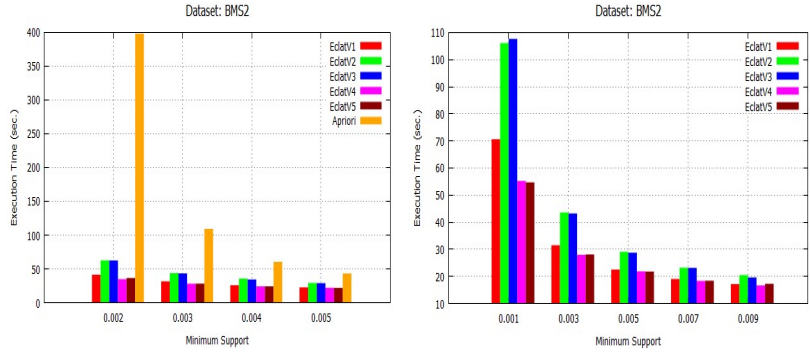

**Fig. 2.** Execution time of algorithms (a) Eclat variants and Apriori (b) Only Eclat variants for varying minimum support on dataset BMS_WebView_2

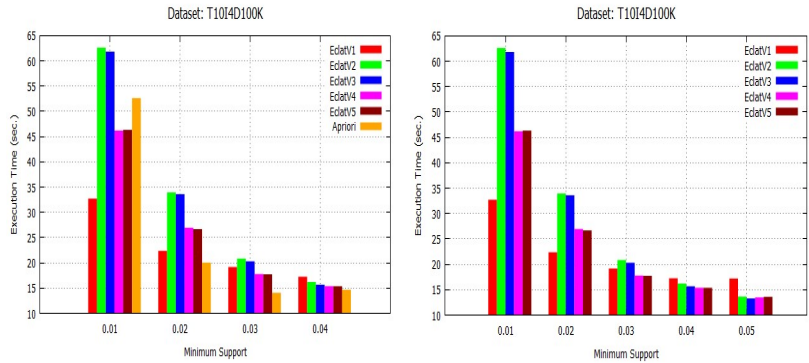

**Fig. 3.** Execution time of algorithms (a) Eclat variants and Apriori (b) Only Eclat variants for varying minimum support on dataset T10I4D100K

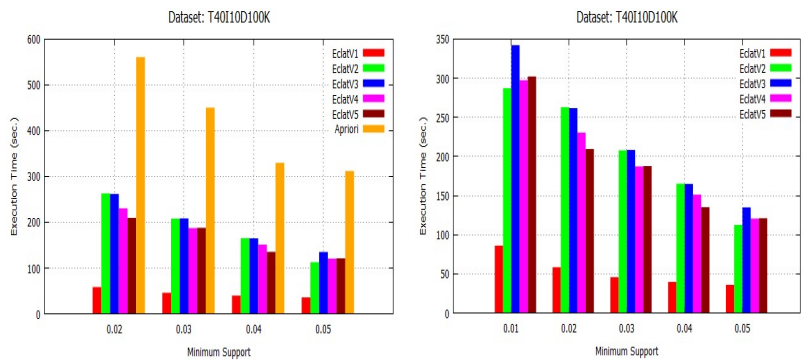

**Fig. 4.** Execution time of algorithms (a) Eclat variants and Apriori (b) Only Eclat variants for varying minimum support on dataset T40I10D100K

Since, the Apriori algorithm is outperformed by all the proposed algorithms, all the subsequent observations are considered only for the proposed algorithms. Figs. 1(b)-4(b) closely compare the execution time of the proposed algorithms EclatV1, EclatV2, EclatV3, EclatV4, and EclatV5. The major algorithmic difference between EclatV1 and EclatV2, EclatV3 is the use of filtered transaction technique in EclatV2 and EclatV3; and the difference between EclatV2, EclatV3 and EclatV4, EclatV5 is the use of hash partitioners for the equivalence class partitioning. EclatV2 and EclatV3 perform worse than EclatV1 (Figs. 1(b)-4(b)). Algorithms EclatV2 and EclatV3 can only improve the performance when they significantly reduce the size of the original transactions after applying the filtered transaction technique. If the size of filtered transactions is still near to that of the original transactions, then it only adds overhead, and increases the overall execution time of the algorithms. Adoption of the filtered transaction technique may improve the performance on a dataset of larger scale where the filtered dataset is reduced significantly. Further, it can be seen that algorithms EclatV4 and EclatV5 significantly improve the performance in comparison to EclatV2 and EclatV3 on all datasets (Figs. 1(b)-4(b)). It proves the effectiveness of equivalence class partitioners used in EclatV4 and EclatV5.

### 5.2 Execution Time on Increasing Number of Executor Cores

The behavior of the proposed algorithms is investigated on the different datasets for the increasing number of executor cores, as shown in Fig. 5(a-b). Execution time has been measured using 2, 4, 6, 8, and 10 executor cores for the two datasets. With the increasing number of cores, execution time of the algorithms decreases. The decline is more apparent in case of BMS2 dataset. It indicates that execution time can be reduced or maintained by allocating more cores or by adding more nodes.

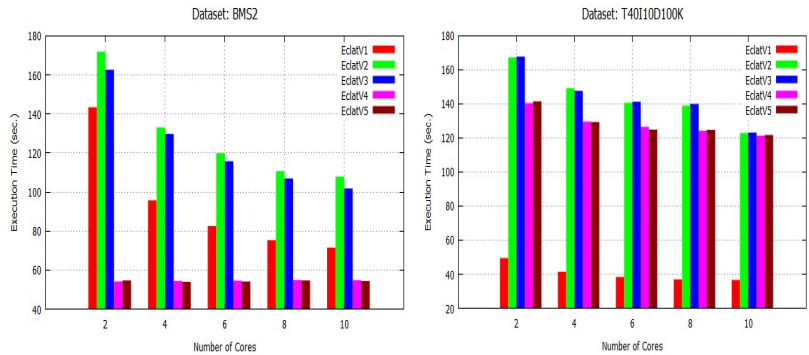

**Fig. 5.** Execution time on varying number of executor cores for two datasets (a) Dataset BMS_WebView_2 at min_sup = 0.001 (b) Dataset T40I10D100K at min_sup = 0.01

### 5.3 Scalability on Increasing Size of Dataset

Scalability test is carried out for the proposed algorithms on the increasing size of dataset T10I4D100K at a fixed value of minimum support, 0.05. To get the larger dataset size, it is doubled each time from its previous dataset, so it ranges from 100K transactions to 1600K transactions as shown in Fig. 6. It can be seen that with the increasing dataset size, execution time of all algorithms increases linearly. It shows the ability of the algorithm to handle the growing size of datasets at the fixed resources. Efficiency can be maintained by adding more resources.

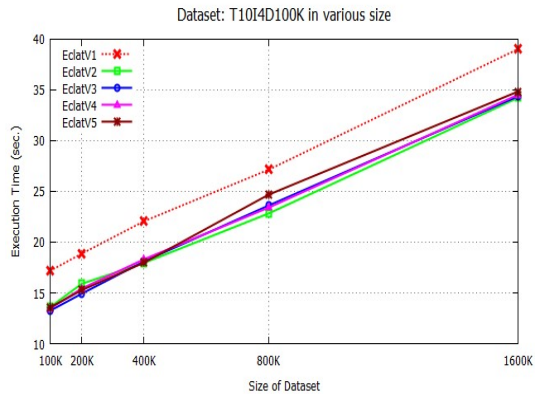

**Fig. 6.** Execution time on increasing size of dataset T10I4D100K at min_sup = 0.05

## 6 Conclusions and Future Directions

The re-designing of Eclat algorithm in the distributed computing environment of Spark has been explored in this paper. The key contribution here is a parallel Eclat algorithm on the Spark RDD framework, named as RDD-Eclat along with the implementation of its five variants. The first variant is EclatV1, and the subsequent variants are EclatV2, EclatV3, EclatV4, and EclatV5. Each variant is resulted from applying some different approach and heuristic on the previous variant. The filtered transaction technique is applied after EclatV1, and the heuristics for equivalence class partitioning are applied in EclatV4 and EclatV5. Experimental results on the both synthetic and real life datasets, shows that all proposed algorithms outperform the YAFIM, a Spark-based Apriori algorithm, by many times, in terms of execution time. As the minimum support threshold decreases, the proposed algorithms perform better in comparison to Spark-based Apriori. Further, the proposed algorithms are closely compared in order to investigate the effect of various heuristics applied on the latter variants. It has been observed that the filtered transaction technique improves the performance when it significantly reduces the size of the original dataset. Further, the heuristics applied in equivalence class partitioning significantly reduce the execution time. Also, the algorithms exhibit scalability when executed on increasing the number of cores and the size of dataset.

Moreover, a more optimized and fine-tuned RDD-Eclat algorithm can be designed in future by efficiently assembling the different modules from the different variants. For example, the heuristic of equivalence class partitioning is not applied in EclatV1 but in EclatV4 and EclatV5 along with the filtered transaction technique. This paper only considers 1-length prefix based equivalence classes, the results can be explored for the k-length prefixes where $k \geq 2$. Also, the heuristic for equivalence class partitioning can be improved further to get a more balanced distribution of equivalence classes.